\documentclass{article}
\usepackage{mrs2005,epsfig}
\begin{document} 
\title{SUB-$L^*$ GALAXIES AT REDSHIFTS z$\sim$4, 3, AND 2: \\ 
THEIR UV LUMINOSITY FUNCTION AND LUMINOSITY DENSITY}

\author{MARCIN SAWICKI}
\affil{Department of Physics, University of California, Santa Barbara, CA 93111, USA\\ {\tt sawicki@physics.ucsb.edu} }

\author{DAVID THOMPSON}
\affil{Caltech Optical Observatories, California Institute of Technology, Pasadena, CA 91125, USA\\ {\tt djt@irastro.caltech.edu} }

\begin{abstract} 
We use very deep (${\cal R}_{lim}$=$27$) $U_nG{\cal R}I$ imaging to
study the evolution of the faint end of the UV-selected galaxy
luminosity function from $z$$\sim$4 to $z$$\sim$2.  We find that the
luminosity function evolves with time and that {\it this evolution is
differential with luminosity}: the number of sub-$L^*$ galaxies
increases from $z$$\sim$4 to $z$$\sim$3 by at least a factor of 2.3,
while the bright end of the LF remains unchanged. Potential systematic
biases restrict our ability to draw strong conclusions at lower
redshifts, $z$$\sim$2, but we {\it can} say that the number density of
sub-$L^*$ galaxies at $z$$\sim$2.2 is {\it at least} as high as it is
at $z$$\sim$3.  Turning to the UV luminosity density of the Universe,
we find that the luminosity density starts dropping with increasing
redshift already beginning at $z$=3 (earlier than recently thought ---
Steidel et al.\ 1999) and that this drop is dominated by the same
sub-$L^*$ galaxies that dominate the evolution of the LF.  This
differential evolution of the luminosity function suggests that
\emph{differentially} comparing key diagnostics of dust, stellar
populations, etc.\ as a function of $z$ and $L$ should let us isolate
the key mechanisms that drive galaxy evolution at high redshift.
\end{abstract} 
 
\section{Introduction} 

The shape of the galaxy luminosity function (LF) bears the imprint of
galaxy formation and evolutionary processes.  In particular, the
presence of a break at the characteristic Schechter luminosity $L^*$,
suggests that galaxies below $L^*$ differ substantially from those
above it. Because of this, our understanding of how galaxies form and
evolve may profit from studying the evolution of not just the bright
but also the faint members of the galaxy population at high redshift.
To date, the faint end of the luminosity function at high redshift has
only been studied using single deep but small fields that do not
provide sufficient galaxy numbers or insurance against cosmic variance
to give trustworthy LF measurements.  And yet, for any reasonable
faint-end slope of the LF, it is exactly these poorly-studied faint
galaxies that not only dominate the number counts, but also contribute
most of the cosmic luminosity density.  With these facts in mind, we
have set out to study the statistics of faint, sub-$L^*$ galaxies at
high redshift.

\section{The Keck Deep Fields survey} 

We study the global statistics of sub-$L^*$ galaxies at high redshift
using a very deep imaging survey that utilizes the very same
$U_nG{\cal R}I$ filter set and color-color selection techniques as are
used for brighter galaxies in the work of Steidel et al.\ (1999, 2003,
2004).  In contrast to the Steidel et al.\ work, our survey reaches
${\cal R}_{lim}$=27; this is 1.5 magnitudes deeper than Steidel et
al.\ and significantly below $L^*$ at $z$=2--4 (Sawicki \& Thompson,
2005).  These Keck Deep Fields (KDF) were obtained with the LRIS
imaging spectrograph on Keck I and represent a total of 71 hours of
integration.  The KDF have an area of 169 arcmin$^2$ divided into 3
spatially-independent patches.  This arrangement allows us to monitor
the effects of cosmic variance on our results.

Our use of the $U_nG{\cal R}I$ filter set lets us select high-$z$
galaxies using the color-color selection criteria defined and
spectroscopically tested by Steidel et al.\ (1999, 2003, 2004).  We
can thus confidently select sub-$L^*$ star-forming galaxies at high
redshift without the need for spectroscopic characterization of the
sample --- characterization that would be extremely expensive at the
faint magnitudes that interest us.  Moreover, this commonality with
the Steidel et al.\ surveys permits us to robustly combine our data
with their, thereby for the first time consistently spanning a large
range in galaxy luminosity at high redshift.

To ${\cal R}$=27, the KDF contains 427, 1481, 2417, and 2043,
$U_nG{\cal R}I$-selected star-forming galaxies at $z$$\sim$4,
$z$$\sim$3, $z$$\sim$2.2, and $z$$\sim$1.7, respectively, selected
using the Steidel et al.\ (1999, 2003, 2004) color-color selection
criteria (Fig.~1).  A detailed description of the Keck Deep Field
observations, data reductions, and high-$z$ galaxy selection can be
found in Sawicki \& Thompson (2005a).

\begin{figure}  
\vspace*{0.5cm}  
\begin{center}
\epsfig{figure=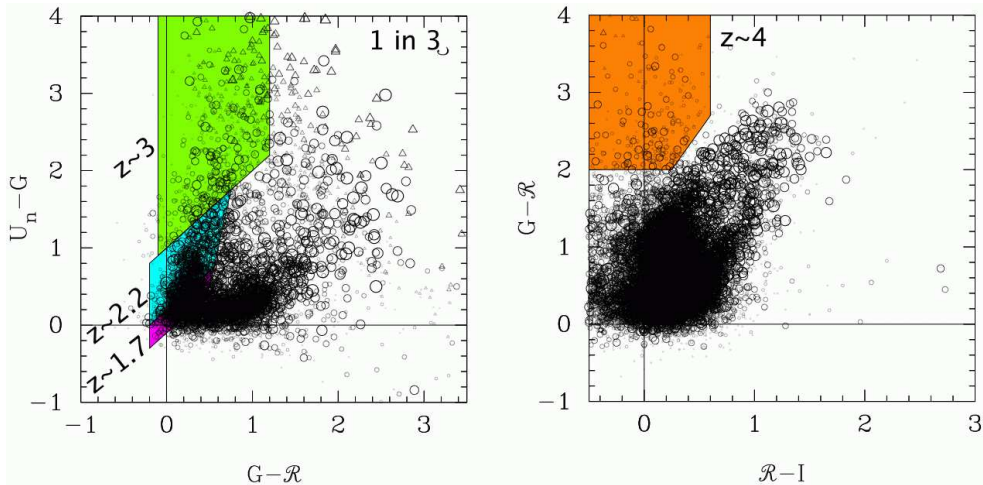,width=13.0cm}  
\end{center}
\vspace*{-0.5cm}
\caption{ Color-color diagrams illustrating the Steidel et al.\ criteria 
we use for selecting high-$z$ samples in the KDF.  The left-hand panel
illustrates selection of $z$$\sim$3, 2.2, and 1.7 samples and the
right-hand panel shows the selection of $z$$\sim$4 galaxies.  For
clarity, in the left-hand panel we plot only one third of the objects
present in our catalog.  The filters we used and our color selection
are {\it both} identical to those used by the Steidel team to select
galaxies at $z$$\sim$1.7--4.  However, our data reach to ${\cal
R}$=27, or 1.5 magnitudes deeper.  }
\end{figure} 

\section{Measuring the luminosity function}

Figure~2(a) shows the luminosity function of UV-selected star-forming
galaxies at $z$$\sim$4, 3, and 2.2.  At $z$$\sim$4 and 3, we augment
our KDF data with the identically-selected Steidel et al.\ (1999)
galaxy count measurements at bright magnitudes, ${\cal R}$$<$25.5.
Following Steidel et al.\ (1999), our LF calculation uses the
effective volume technique, computing $V_{eff}$ through recovery tests
of artificial high-$z$ galaxies implanted into the imaging data.  Our
analysis includes a comprehensive study of a variety of potential
systematic effects (see Sawicki \& Thompson 2005b for details) and
consequently we can be confident about the robustness of our LFs.
When such robustness is lacking --- as it is at $z$$\sim$2 --- we know
not to overinterpret our results.

As Fig.~2(a) shows, our data reach very faint luminosities which
correspond to star formation rates of only 1--2~$M_\odot$/yr (not
corrected for dust).  Until now, LF measurements at these depths have
only been carried out in small, single deep fields (such as the HDFs)
and so are subject to small number statistics and/or vagaries of large
scale structure (e.g., Sawicki, Lin, \& Yee 1997; Gabasch et al.\
2004).  Moreover, until now it was necessary to combine these
faint-end data with bright-end results that had been selected using
different selection criteria --- an approach fraught with potential
biases.  And finally, hitherto little attention has been paid to
potential systematic effects in high-z LF measurements. In contrast,
our KDF results are the deepest and most robust LF measurements at
these redshifts and, very importantly, can readily and
straightforwardly be combined with the Steidel et al.\ (1999) results
at the bright end.

Our $z$$\sim$4 and $z$$\sim$3 LFs are very insensitive to a large
range of systematics, although the $z$$\sim$2.2 and $z$$\sim$1.7 LFs
are more uncertain.  Specifically, the $z$$\sim$1.7 LF is likely quite
strongly affected by systematics (see Sawicki \& Thompson 2005b for
details) and so we do not discuss it here.  Our $z$$\sim$2.2 LF
measurement is probably also affected by systematics but to a smaller
degree than the $z$$\sim$1.7 LF: we can confidently state that the
number density of $z$$\sim$2.2 galaxies is {\it not lower} than shown
in Fig.~2(b), and that if it is higher than the LF shown, it is so by
no more than a factor of $\sim$2.  The $z$$\sim$3 and $z$$\sim$4 LFs
{\it are} very robust against systematics as verified by a range of
tests (Sawicki \& Thompson 2005b).

The values of Schechter function parameters are given in Sawicki \&
Thompson (2005b) and here we only note that our $M^*$$\sim$$-$21 at
$z$$\sim$4 and $z$$\sim$3 and also that we find shallower faint-end
slopes than found by Steidel et al.\ (1999) using HDF-N data.
Specifically, Steidel et al.\ (1999) found $\alpha$=$-$1.6 at
$z$$\sim$3 in the HDF-N (and in the absence of good statistics, had to
{\it assume} the same $\alpha$=$-$1.6 value for $z$$\sim$4).  Our KDF
analysis, which uses much larger datasets that span several spatially
independent fields, finds $\alpha$=$-$1.43 at $z$$\sim$3 and
$\alpha$=$-$1.26 at $z$$\sim$4.  Workers who still use the Steidel et
al.\ (1999) HDF-N value to extrapolate the contributions of sub-$L^*$
galaxies from bright galaxy counts at higher redshifts should take
note of these more accurate, shallower faint-end slopes.  The new,
shallower faint-end slopes we find have a significant effect on
estimates of the luminosity density of the Universe (see \S~5).

\begin{figure}  
\vspace*{0.25cm}  
\begin{center}
\epsfig{figure=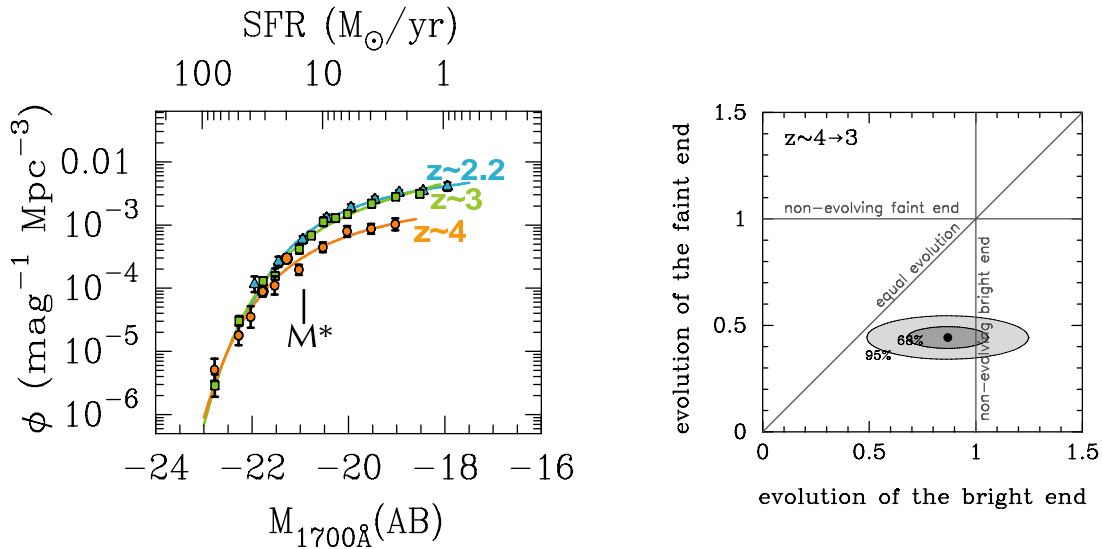,width=14.5cm}  
\end{center}
\vspace*{-0.5cm}  
\caption{{\bf (a) Left panel:} Evolution of the UV luminosity function.  
The $z$$\sim$4 and $z$$\sim$3 LFs are robust against a range of
possible systematics and the evolution between these two epochs is
real. The $z$$\sim$2.2 LF may be subject to a systematic bias --- the
$z$$\sim$2.2 LF shown here is the {\it lower limit} on the number
density of galaxies.  At $z$$\sim$4 and $z$$\sim$3 the KDF data are
complemented by data from Steidel et al.\ (1999) at the bright end
($M$$<$$M^*$ at $z$$\sim$4 and $M$$<$$M^*$$+$1 at $z$$\sim$3).  The
errorbars include both root-N statistics and field-to-field variance
estimated through bootstrap resampling.  {\bf (b) Right panel:} The
luminosity dependent evolution of the number density of galaxies from
$z$$\sim$4 to $z$$\sim$3. The horizontal axis shows the change in the
number density of luminous galaxies ($M_{1700}$$<$$-$21) between
$z$$\sim$4 and $z$$\sim$3. The vertical axis shows the evolution for
low-luminosity galaxies, ($M_{1700}$$>$$-$21).  Three fiducial
evolutionary scenarios are shown using solid lines: no faint-end
evolution (horizontal line), no bright-end evolution (vertical line)
and equal evolution at the bright and faint ends (diagonal
line). There is substantial evolution of the low-luminosity population
from $z$$\sim$4 to $z$$\sim$3 and the differential evolution scenario
is statistically significant at the 98.5\% level. }
\end{figure} 

\section{Differential evolution of the luminosity function}

The luminosity function of high-$z$ galaxies undergoes evolution that
is differential with luminosity. As Fig.~2 shows, we find a factor of
$\sim$2.5 evolution in the number density (or, alternatively, 2 mag in
luminosity) of \emph{faint}, sub-$L^*$ Lyman Break Galaxies between
$z$$\sim$4 to $z$$\sim$3.  At the same time, there is no evidence for
evolution of the bright end of the LF.  

Our tests show that it is highly unlikely that the observed
differential evolution is due to some systematic effect such as
selection bias, our modeling of the survey volume, etc.\ (see Sawicki
\& Thompson 2005b for a detailed discussion).  The effect is also 
statistically quite significant --- the probability that the evolution
is not differential with luminosity is only 1.5\% (see Fig.~2(b)).  We
therefore conclude that the differential, luminosity-dependent
evolution is very likely real.  If so, it must reflect some real
intrinsic evolutionary differences between faint and luminous LBGs.
Understanding what these differences are will give us insights into
what drives galaxy evolution at high redshift.

\section{The UV luminosity density of the Universe} 

The importance of the faint end of the luminosity function extends
into measurements of the cosmic luminosity density and its derivative,
the cosmic star formation rate.  As is illustrated in Fig.~3(a), for
observed values of the faint-end slope $\alpha$ the bulk of the
luminosity density resides in galaxies that are fainter than $L^*$.
Attempts to measure the contribution of the low-luminosity galaxies
have been made previously using single small fields (e.g., Madau et
al.\ 1996; Sawicki, Lin, \& Yee, 1997; Steidel et al.\ 1999; Gabash et
al 2004).  But it is only with a deep, mutli-field and large-area
survey such as the KDF that we can robustly measure the contribution
of these dominant galaxies to the total luminosity density.

Figure 3(b) shows a compilation of UV luminosity density values that
trace the evolution of the luminosity density from $z$$\sim$5 to the
present.  The black symbols show the total luminosity density,
obtained by integrating the luminosity-weighted luminosity function
over {\it all} luminosities.  The total luminosity density exhibits a
strong increase from $z$$\sim$5 to $z$$\sim$3 followed by the
well-known strong decline below $z$$\sim$1.  At present, it is not yet
clear if there is a plateau at $z$$\sim$3--2 since, as mentioned in
\S~3, our $z$$\sim$2 points are strictly speaking lower limits.  

It is interesting to ask which galaxies contain the bulk of the UV
light and --- presumably --- star formation at each redshift.  The
colored points in Fig.~3(b) show the contributions to the the total
luminosity density that are made by galaxies in three luminosity
ranges (split at $M$=$-$21 and $M$=$-$18).  As the red diamonds show,
the {\it bulk} of the luminosity density at high redshift comes from
intermediate-luminosity galaxies between $M^*_{z\sim3,4}$ and
$M^*_{z\sim3,4}$$+$3. Galaxies that are brighter than $M$=$-$21
($\approx$$M^*$) as well as those fainter than $M$=$-$18
($\approx$$M^*$$+$3) contribute little to the total luminosity density
at high redshift.  The galaxies that do dominate the luminosity
density come from just a small range of luminosity just below $L^*$
and are exactly the galaxies probed by the KDF.

\begin{figure}  
\vspace*{0.25cm}  
\begin{center}
\epsfig{figure=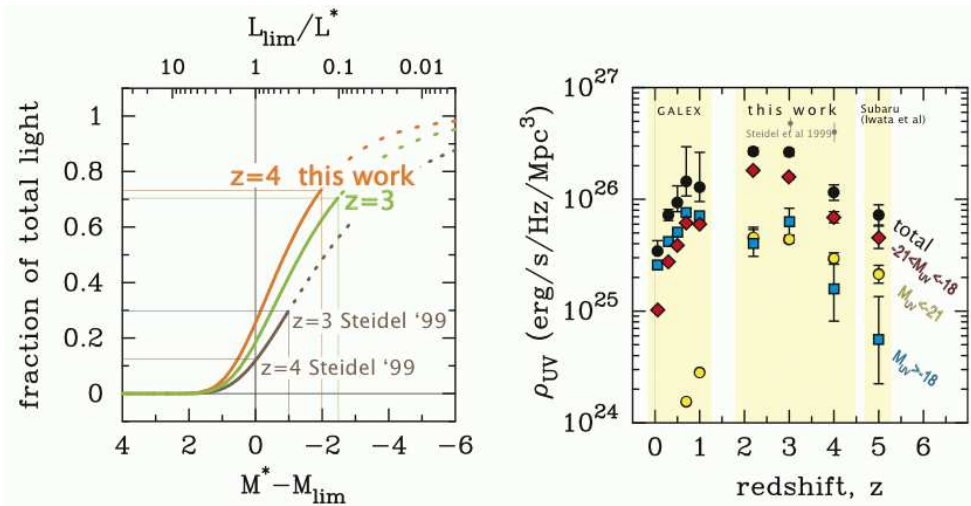,width=13.0cm}  
\end{center}
\vspace*{-0.5cm}  
\caption{{\bf (a) Left panel:} The fraction of total light captured by 
integrating the luminosity function down to a limiting magnitude
$M_{lim}$.  The curves are computed by integrating the
luminosity-weighted Schechter function with parameters derived by
Steidel et al.\ (1999; blue curves) and by our analysis (orange and
green curves).  Steidel et al.\ ground-based surveys capture only a
small fraction of the light produced by Lyman Break Galaxies.  In
contrast, the KDF are sufficiently deep to capture 75-80\% of the UV
light present at these redshifts.  {\bf (b) Right panel:} The UV
luminosity density of the Universe.  Solid black symbols show the
total luminosity density, while the colored symbols show the
luminosity density that resides in galaxies split by luminosity
range. The low-z points are due to GALEX (Arnouts et al.\ 2005;
Schiminovich et al.\ 2005; Wyder et al.\ 2005) where we used their
results to calculate values for our three luminosity sub-ranges; we
calculated the $z$$\sim$5 points by integrating the LF of Iwata et
al.\ (2005) based on deep Subaru imaging.  The values obtained by
integrating the Steidel et al.\ (1999) LFs (with their
$\alpha$$=$$-$1.6) are also shown for comparison. This figure shows
that the bulk of the UV luminosity density at high redshifts comes
from galaxies in the range $-$21$<$$M$$<$$-$18 (red diamonds).  These
are exactly the sub-$L^*$ galaxies that are probed by the KDF and are
also the galaxies responsible for the rapid rise of the luminosity
density from $z$$\sim$5 to $z$$\sim$3.}
\end{figure} 

\section{What drives the evolution of the sub-$L^*$ galaxies?} 

The steep increase in the luminosity density at high redshift is due
to the increase in the number density of sub-$L^*$ galaxies and the
resulting steepening of the LF's faint-end slope $\alpha$.  However,
as Fig.~2 illustrates, while the number density of sub-$L^*$ galaxies
increases with time, the number of luminous ones remains constant.
This differential evolution must reflect important evolutionary
differences between luminous and faint galaxies at these redshifts.
Understanding the mechanism that underlies this differential,
luminosity-dependent evolution will teach us important lessons about
how galaxies form and evolve.

At present, it is not clear what is responsible for the observed
differential evolution of the LF.  One straightforward possibility is
that the number of faint (but not bright) LBGs simply increases over
the 600~Myr from $z$$\sim$4 to $z$$\sim$3. If UV luminosity reflects
underlying mass, such an increase could be a reflection of an
evolutionary trend that favors star formation in progressively
lower-mass systems as the Universe ages (see, e.g., Iwata et al.\
2005). Another possibility is that dust properties --- its amount or
covering fraction --- may be decreasing in faint LBGs with time.  Such
a change would make individual faint galaxies brigher and thus would
steepen the faint end of the LF.  Yet another possibility is that if
star formation in individual faint LBGs is episodic (e.g., Sawicki \&
Yee 1998), then they may brighten and fade (and thus move in and out
of a given magnitude bin in the LF) with duty cycle properties that
change with redshift.  Longer or more frequent star-forming episodes
would then result in a steeper luminosity function.

Whatever the mechanism, the fact that the LF evolution is differential
suggests that different evolutionary mechanisms are at play as a
function of UV luminosity. Studies that {\it differentially} compare
key galaxy properties as a function of luminosity and redshift should
help us isolate the mechanisms that are responsible for this
evolution.  Such studies, including comparisons between spectral
energy distributions or clustering properties of galaxies as a
function of luminosity are already underway.  A key point here is that
we have identified luminosity and redshift as important variables in
galaxy evolution at high $z$.  While LBG follow-up studies to date
have primarily focused on luminous galaxies, extending such studies to
cover a range of luminosity that spans the break ``knee'' of the
luminosity function will yield valuable insights into how galaxies
form and evolve.

\vfill 

\begin{thebibliography}{}{ 
\bibitem{} Arnouts, S., et al.\ 2005, ApJ, 613, L43 
\bibitem{} Gabasch, A., et al.\ 2004, A\&A, 421, 41
\bibitem{} Iwata, I., et al.\ 2005, these proceedings
\bibitem{} Madau, P. et al.\ 1996, MNRAS, 293, 1388
\bibitem{} Sawicki, M.J., Lin, H., \& Yee, H.K.C. 1997, 113, 1 
\bibitem{} Sawicki, M., \& Thompson, D. 2005a, ApJ, in press, astro-ph/0507424
\bibitem{} Sawicki, M., \& Thompson, D. 2005b, ApJ, submitted, astro-ph/0507519
\bibitem{} Sawicki, M. \& Yee, H.K.C. 1998, AJ, 115, 1329
\bibitem{} Schiminovich, D., et al.\ ApJ, 619, L47
\bibitem{} Steidel, C.C., et al.\ 1999, ApJ, 519, 1
\bibitem{} Steidel, C.C., et al.\ 2003, ApJ, 592, 728
\bibitem{} Steidel, C.C., et al.\ 2004, ApJ, 604, 534
\bibitem{} Wyder, T., et al.\ 2005, ApJ, 619, L15
}
\end{thebibliography}
\end{document}